\documentclass[12pt]{article}
\begin{document}
\begin{titlepage}
\begin{flushright}
hep-th/0508052\\
TIT/HEP-541\\
August, 2005\\
\end{flushright}
\vspace{0.5cm}
\begin{center}
{\Large \bf 
Non(anti)commutative ${\cal N}=2$ Supersymmetric\\
$U(N)$ Gauge Theory and\\ Deformed Instanton Equations
}
\lineskip .75em
\vskip2.5cm
{\large Katsushi Ito and Hiroaki Nakajima}
\vskip 2.5em
{\large\it Department of Physics\\
Tokyo Institute of Technology\\
Tokyo, 152-8551, Japan}  \vskip 4.5em
\end{center}
\begin{abstract}
We study deformed supersymmetry in ${\cal N}=2$ supersymmetric 
$U(N)$ gauge theory in non(anti)commutative ${\cal N}=1$ superspace. 
Using the component formalism, 
we construct deformed ${\cal N}=(1,1/2)$ supersymmetry explicitly. 
Based on the deformed supersymmetry, 
we discuss the $C$-dependence of the correlators.
We also study the $C$-deformation of the instanton equation 
for the gauge group $U(2)$.

\end{abstract}
\end{titlepage}

\baselineskip=0.7cm
Supersymmetric field theories in non(anti)commutative superspace 
\cite{ScNi,Se}
has been  attracted much interests from  the viewpoint of 
effective field theories on D-branes in the graviphoton
background \cite{OoVa,BeSe,DeGrNi}.
Superstrings in this  background provide 
some interesting low-energy physics in ${\cal N}=2$
supersymmetric field theories \cite{Ne}
and their ${\cal N}=1$ deformations \cite{DiVa}.
It would be important to study
${\cal N}=2$ supersymmetric gauge theories in non(anti)commutative 
superspace in order to understand  graviphoton effects in 
the low-energy
effective theories from the microscopic point of view.

It is convenient to use ${\cal N}=2$ extended
non(anti)commutative  superspace for studying
non(anti)commutative gauge theories
 where supersymmetry is manifestly realized
\cite{KlPeTa,FeSo,IvLeZu,ArItOh2,KeSa,IvZu,FeIvLeSoZu}.
In previous papers \cite{ArItOh2,ArItOh4,ArItOh5}, 
${\cal N}=2$ supersymmetric $U(1)$ gauge
theory in non(anti)commutative ${\cal N}=2$ harmonic superspace has
been studied.
We have computed the deformed Lagrangian up to the first order in the
deformation parameter $C$ of the superspace and examined their deformed
symmetries.
It is, however, difficult to calculate higher order 
$C$-corrections and extend the $U(1)$ gauge group to $U(N)$.

There exist two cases such that the deformed Lagrangian 
of ${\cal N}=2$ supersymmetric $U(N)$ gauge theory becomes simple.
One is the case of the singlet deformation where the deformation
parameter
belongs to the singlet representation of the  $R$-symmetry group
$SU(2)$ \cite{FeSo,IvLeZu,ArIt3,IvZu,FeIvLeSoZu}.
The other is the case that one introduces only 
deformation into ${\cal N}=1$ subsuperspace of ${\cal N}=2$ superspace.
In a recent paper \cite{ArItOh5}, it is shown that 
the $O(C)$ Lagrangian of the $U(1)$ theory defined in 
non(anti)commutative ${\cal N}=2$ harmonic superspace 
leads to the theory in the non(anti)commutative ${\cal N}=1$ superspace
\cite{Se} by
the reduction of deformation parameters and 
some field redefinitions.
It is also shown that 
the theory has ${\cal N}=(1,1/2)$ supersymmetry 
consistent with the Poisson structure of the theory.
Here ${\cal N}=(1,1/2)$ means that there are two chiral 
and one antichiral supercharges,  as in \cite{IvLeZu}.

In this paper, we will 
study  ${\cal N}=2$ supersymmetric $U(N)$ gauge theory
in the deformed ${\cal N}=1$ superspace, whose
Lagrangian has been constructed in \cite{ArItOh1}.
We will construct deformed ${\cal N}=(1,1/2)$ supersymmetry explicitly.
Based on this symmetry, we will study the $C$-deformed correlators of 
the observables. 
We will also examine the the $C$-deformed instanton equations 
for the gauge group $U(2)$. 
The instanton solutions in non(anti)commutative ${\cal N}=1$ gauge 
theory have been investigated in \cite{Im,inst}. 
In the case of the gauge group $U(2)$\cite{Im}, 
it has been found that the $SU(2)$ part of 
the instanton equations is not deformed and 
the $U(1)$ part is deformed only. 
In the non(anti)commutative ${\cal N}=2$ $U(2)$ theory, 
we will show that the $SU(2)$ part of the
instanton equations is not deformed except for one of the fermions.

Let $(x^m,\theta^{\alpha},\bar{\theta}^{\dot{\alpha}})$ 
($m=0,\ldots,3$, $\alpha,\dot{\alpha}=1,2$) be 
supercoordinates of ${\cal N}=1$  superspace 
and $\sigma^m_{\alpha\dot{\alpha}}$ and 
$\bar{\sigma}^{m\dot{\alpha}\alpha}$ Dirac matrices.
We will study Euclidean spacetime so that chiral and antichiral 
fermions transform independently under the Lorentz transformations.
$Q_\alpha={\partial\over \partial\theta^\alpha}
-i\sigma^{m}_{\alpha\dot{\alpha}}
\bar{\theta}^{\dot{\alpha}}\partial_m$ and
$\bar{Q}^{\dot{\alpha}}=
-{\partial\over \partial\bar{\theta}_{\dot{\alpha}}}+i
\theta_{\alpha}\bar{\sigma}^{m\dot{\alpha}\alpha}
\partial_m$ are supercharges.
$D_\alpha={\partial\over \partial\theta^\alpha}
+i\sigma^{m}_{\alpha\dot{\alpha}}
\bar{\theta}^{\dot{\alpha}}
\partial_m$
and
$\bar{D}_{\dot{\alpha}}=
-{\partial\over \partial\bar{\theta}_{\dot{\alpha}}}-i
\theta_{\alpha}\bar{\sigma}^{m\dot{\alpha}\alpha}
\partial_m$ are the supercovariant derivatives.
$\sigma^{mn}={1\over4}(\sigma^m\bar{\sigma}^n
-\sigma^n\bar{\sigma}^m)$,
and 
$\bar{\sigma}^{mn}={1\over4}(\bar{\sigma}^m\sigma^n
-\bar{\sigma}^n\sigma^m)$
are the Lorentz generators.
Here we will follow the conventions of Wess and Bagger \cite{WeBa}.

The non(anti)commutativity in ${\cal N}=1$ superspace is introduced by 
the $*$-product:
\begin{equation}
 f*g(x,\theta,\bar{\theta})=f(x,\theta,\bar{\theta})
\exp\left(-{1\over2}\overleftarrow{Q}_\alpha C^{\alpha\beta}
\overrightarrow{Q}_\beta \right)g(x,\theta,\bar{\theta}).
\end{equation}
Using this $*$-product, the anticommutation relations for 
$\theta$ become
\begin{eqnarray}
 \left\{\theta^\alpha, \theta^\beta\right\}_{*}&=& C^{\alpha\beta}
\end{eqnarray}
while the chiral coordinates $y^m=x^m+i\theta\sigma^m\bar{\theta}$ and
$\bar{\theta}$ are still commuting and anticommuting coordinates, 
respectively.

${\cal N}=2$ supersymmetric $U(N)$ gauge theory in this deformed
superspace was formulated in \cite{ArItOh1}.
It can be constructed by 
vector superfields $V$, chiral superfields $\Phi$ and an anti-chiral
superfields
$\bar{\Phi}$, where $\Phi$ and $\bar{\Phi}$ belong to the adjoint 
representation of $U(N)$.
We introduce the basis $t^a$ ($a=1,\cdots, N^2$) of Lie algebra of $U(N)$,
normalized as ${\rm tr}(t^a t^b)=k\delta^{ab}$. 
The Lagrangian is 
\begin{eqnarray}
{\cal L}
={1\over k}
\int d^2\theta d^2\bar{\theta}\,
{\rm tr}(\bar{\Phi}*e^{V}*\Phi* e^{-V})
+{1\over 16kg^2}{\rm tr}\left(\int d^2\theta
 W^{\alpha}*W_{\alpha}
+\int d^2\bar{\theta}\bar{W}_{\dot{\alpha}}*\bar{W}^{\dot{\alpha}}
\right)
\label{eq:lag1}
\end{eqnarray}
where $g$ denotes the coupling constant.
$W_\alpha = -{1\over4}\bar{D}^2e^{-V}D_\alpha e^{V}$ 
and $\bar{W}_{\dot{\alpha}}={1\over 4}D^2
e^{-V}\bar{D}_{\dot{\alpha}}e^{V}$
are the chiral and antichiral field strengths.
Note that multiplication of superfields are defined by the $*$-product.

This Lagrangian is invariant under the gauge transformations
$\Phi\rightarrow e^{-i\Lambda}*\Phi* e^{i\Lambda}$, 
$\bar{\Phi}\rightarrow e^{-i\bar{\Lambda}}*\bar{\Phi}*e^{i\bar{\Lambda}}$
and $e^{V}\rightarrow e^{-i\bar{\Lambda}}*e^{V}*e^{i\Lambda}$.
To write down the Lagrangian in terms of component fields, 
it is convenient to take the Wess-Zumino(WZ) gauge as in the commutative case.
Since the $*$-product deforms the gauge transformation,
it is necessary to redefine the component fields such that 
these transform canonically under the gauge
transformation\cite{Se,ArItOh1}.
For ${\cal N}=2$ $U(N)$ theory, these superfields in the WZ gauge are 
\begin{eqnarray}
 \Phi(y,\theta)&=& A(y)+\sqrt{2}\theta\psi(y)+\theta\theta F(y),
\nonumber\\
\bar{\Phi}(\bar{y},\bar{\theta})&=&
\bar{A}(\bar{y})+\sqrt{2}\bar{\theta}\bar{\psi}(\bar{y})
+\bar{\theta}\bar{\theta}\left(
\bar{F}+i C^{m n}\partial_m\left\{ v_{n},\bar{A}\right\}
-{1\over4}C^{mn}\left[ v_{m},\left\{ v_{n},\bar{A}\right\}\right]
\right)(\bar{y}),
\nonumber\\
 V(y,\theta,\bar{\theta})&=&
-\theta \sigma^{\mu}\bar{\theta} v_{\mu}(y)
+i \theta\theta \bar{\theta}\bar{\lambda}(y)
-i \bar{\theta}\bar{\theta}\theta^{\alpha}
\left(\lambda_\alpha
+{1\over4}\varepsilon_{\alpha\beta}C^{\beta\gamma}
\left\{(\sigma^{\mu}\bar{\lambda})_{\gamma}, v_{\mu}\right\}
\right)(y)
\nonumber\\
&& +{1\over2} \theta\theta \bar{\theta}\bar{\theta}
(D-i\partial^{\mu}v_{\mu})(y).
\label{eq:wz1}
\end{eqnarray}
Here $\bar{y}^m=x^m-i\theta\sigma^m\bar{\theta}$ are the antichiral
coordinates
and
$C^{mn}=C^{\alpha\beta}\varepsilon_{\beta\gamma}(\sigma^{mn})_{\alpha}
{}^{\gamma}$.
Since $\sigma^{mn}$ is self-dual, $C^{mn}$ is also self-dual.
Substituting (\ref{eq:wz1}) into the Lagrangian (\ref{eq:lag1}), we
obtain
the deformed Lagrangian written in
terms of component fields.
In this expression, however, normalizations of 
two fermions $\psi$ and $\lambda$ are different.
In order to see symmetries between two fermions manifestly,
it is useful to rescale $V$ to $2g V$ and $C^{\alpha\beta}$ to ${1\over
2g}C^{\alpha\beta}$. 
Then the Lagrangian takes the form 
 ${\cal L}={\cal L}_0+{\cal L}_1$. 
Here ${\cal L}_0$ is the undeformed Lagrangian with the topological term:
\begin{eqnarray}
 {\cal L}_0&=&
\frac{1}{k}{\rm tr}\Bigl(
-\frac{1}{4}F^{m n}F_{m n}-\frac{1}{4}F^{m n}\tilde{F}_{m n}
-i \bar{\lambda}\bar{\sigma}^{m}D_{m}\lambda
+\frac{1}{2}\tilde{D}^2\nonumber\\
&&
\!\!\!\!\!\!\!\!\!\!\!\!
-(D^{m}\bar{A})D_{m}A
-i \bar{\psi}\bar{\sigma}^{m}D_{m}\psi
+\bar{F} F
-i\sqrt{2} g[\bar{A},\psi]\lambda
-i\sqrt{2} g[A,\bar{\psi}]\bar{\lambda}
-{g^2\over2}[A,\bar{A}]^2
\Bigr),
\label{eq:lag2a}
\end{eqnarray}
where $F_{mn}=\partial_m v_n-\partial_n v_m+ig [v_m,v_n]$, 
$\tilde{F}_{mn}=\frac{1}{2}\epsilon_{mnpq}F^{pq}$ and
$D_m \lambda=\partial_m \lambda+ig [v_m, \lambda]$ etc.
We have also introduced an auxiliary field $\tilde{D}$ defined by
$\tilde{D}=D+g [A,\bar{A}]$ in order to see undeformed ${\cal N}=2$
supersymmetry in a symmetric way.
${\cal L}_1$ is the $C$-dependent part of the Lagrangian:
\begin{eqnarray}
{\cal L}_1 &=&
\frac{1}{k} {\rm tr} 
\left( - \frac{i}{2} C^{m n} F_{m n} \bar{\lambda} 
\bar{\lambda} 
+ \frac{1}{8} |C|^2 (\bar{\lambda} \bar{\lambda})^2 \right. 
\nonumber\\
& & \left. +\frac{i}{2} C^{m n} F_{m n} \{ \bar{A},F \} 
- \frac{\sqrt{2}}{2} C^{\alpha \beta} \{ D_{m} \bar{A} , 
(\sigma^{m} \bar{\lambda})_{\alpha} \} \psi_{\beta}  
- \frac{1}{16} |C|^2 [ \bar{A} , \bar{\lambda}] [\bar{\lambda} , F] 
\right). 
\label{eq:lag2b}
\end{eqnarray}
Here $|C|^2=C^{mn}C_{mn}$.

In the case of $C=0$, 
the action is invariant under ${\cal N}=2$ supersymmetry
transformations, where only ${\cal N}=1$ supersymmetry 
generated by $Q_\alpha$ and $\bar{Q}^{\dot{\alpha}}$ are manifestly
realized in ${\cal N}=1$ superspace.
Other ${\cal N}=1$ supersymmetry would be realized manifestly when we use 
${\cal N}=2$ extended superspace. In particular ${\cal N}=2$ harmonic
superspace \cite{GaIvOgSo} provides very efficient tools to study off-shell 
${\cal N}=2$ supersymmetric field theories.
The most general non(anti)commutative deformations are
studied by using extended superspace.
We expect that this deformed theory 
is derived from the non(anti)commutative ${\cal N}=2$ harmonic
superspace by the reduction of deformation parameters, which will be 
discussed in a separate paper.

In previous papers\cite{ArItOh2,ArItOh4,ArItOh5}, we have studied
${\cal N}=2$ supersymmetric $U(1)$ gauge theory and their deformed
supersymmetry structure in the component formalism.
In particular, for generic deformation, ${\cal
N}=(1,0)$ deformed supersymmetry has been constructed up to the first order 
of the deformation parameters.
When the deformation parameters are reduced such that only ${\cal N}=1$
subspace becomes non(anti)commutative, 
the deformed supersymmetry is enhanced to ${\cal N}=(1,1/2)$ supersymmetry. 
This is because
supersymmetries other than 
 $\bar{Q}^{\dot{\alpha}}$ is consistent with the Poisson structure
of the deformed superspace \cite{IvLeZu}. 
In this reduced case, it is shown that the $O(C)$ Lagrangian defined 
in the ${\cal N}=2$ harmonic superspace is equal to that of the deformed
${\cal N}=1$ superspace by the field redefinitions.

We now study the $U(N)$ case.
The undeformed superfield action is invariant under 
${\cal N}=1$ supersymmetry generated by
$\xi Q+\bar{\xi}\bar{Q}$. 
Since this transformation does not preserve the WZ gauge, 
we need to do gauge transformation to retain the WZ gauge.
Then the (undeformed) 
supersymmetry transformations $\delta^0_\xi$ and $\delta^0_{\bar{\xi}}$ 
of the component fields in the 
WZ gauge are 
\begin{eqnarray}
\delta^0_{\xi}v_m&=& i\xi\sigma_m\bar{\lambda},
\nonumber\\
\delta^0_{\xi} \lambda&=&
i\xi \tilde{D}-ig \xi[A,\bar{A}]+\sigma^{mn}\xi F_{mn},
\quad
\delta^0_{\xi}\bar{\lambda}=0, 
\nonumber\\
\delta^0_{\xi}\tilde{D}&=&
-\xi 
\sigma^m D_m\bar{\lambda}
+\sqrt{2}g[\xi\psi, \bar{A}], \nonumber\\
 \delta^0_{\xi}A&=&\sqrt{2}\xi\psi,\quad
 \delta^0_{\xi}\psi=\sqrt{2}\xi F,\quad
\delta^0_{\xi}F=0, \nonumber\\
 \delta^0_{\xi}\bar{A}&=&0, \quad
 \delta^0_{\xi}\bar{\psi}=
\sqrt{2}i \bar{\sigma}^m \xi D_m \bar{A}, \quad
 \delta^0_{\xi}\bar{F}=
i\sqrt{2}\xi\sigma^{m}D_{m}\bar{\psi}-2g i\xi[\bar{A},\lambda],
\label{eq:susy0}
\end{eqnarray}
\begin{eqnarray}
\delta^0_{\bar{\xi}}v_m&=& 
i\bar{\xi}\bar{\sigma}_m\lambda,
\nonumber\\
\delta^0_{\bar{\xi}} \lambda&=&
0,
\quad
\delta^0_{\bar{\xi}}\bar{\lambda}=-i\bar{\xi}\tilde{D}+ig \bar{\xi}[A,\bar{A}]
+\bar{\sigma}^{mn}\bar{\xi}F_{mn} ,
\nonumber\\
\delta^0_{\bar{\xi}}\tilde{D}&=&
\bar{\xi}\bar{\sigma}^n D_n \lambda
+\sqrt{2}g [A,\bar{\xi}\bar{\psi}], \nonumber\\
 \delta^0_{\bar{\xi}}A&=&0,\quad
 \delta^0_{\bar{\xi}}\psi=
\sqrt{2}i \sigma^m\bar{\xi}D_m A, \quad
\delta^0_{\bar{\xi}}F=\sqrt{2}i\bar{\xi}\bar{\sigma}^mD_m\psi
+2g i\bar{\xi}[\bar{\lambda},A],\nonumber\\
 \delta^0_{\bar{\xi}}\bar{A}&=&\sqrt{2}\bar{\xi}\bar{\psi},\quad
 \delta^0_{\bar{\xi}}\bar{\psi}=\sqrt{2} \bar{\xi}\bar{F},
\quad
 \delta^0_{\bar{\xi}}\bar{F}=
0.
\label{eq:susy0b}
\end{eqnarray}
The remaining ${\cal N}=1$ supersymmetry denoted by
$\delta^0_\eta$ and $\delta^0_{\bar{\eta}} $ 
can be obtained from (\ref{eq:susy0}) by using the $R$-symmetry:
$
 \xi\rightarrow \eta,\quad
 \lambda\rightarrow -\psi, \quad
 \psi\rightarrow \lambda,\quad
\tilde{D}\rightarrow -\tilde{D},\quad
F\rightarrow \bar{F}.
$

Now we will construct the deformed ${\cal N}=(1,1/2)$ supersymmetry
which keeps the $U(N)$ Lagrangian ${\cal L}$ invariant up to the total
derivatives.
The term ${\cal L}_1$ is not invariant under the undeformed
supersymmetry transformations $\delta^0_\xi$, $\delta^0_\eta$ and
$\delta^0_{\bar{\eta}}$.
Since the deformed term ${\cal L}_1$ is a polynomial in $C$, 
we denote ${\cal L}^{(n)}_1$ ($n\geq 1$) by its $n$-th order term in $C$.
The deformed supersymmetry transformations can be expanded in the form
$\delta=\delta^0+\delta^1+\cdots$.
Here $\delta^n$ is the $n$-th order term in $C$.
$\delta^n$ is determined recursively by solving the conditions
$\delta^1{\cal L}_0+\delta^0{\cal L}^{(1)}_1=0$ and
$\delta^2{\cal L}_0+\delta^1{\cal L}^{(1)}_1+\delta^0 {\cal
L}^{(1)}_2=0$ and so on.

The deformed transformation $\delta_\xi$, which was calculated
 in \cite{ArItOh1}, takes the form
$\delta_{\xi}=\delta^0_\xi+\delta^1_{\xi}$ and is given by
\begin{eqnarray}
 \delta_{\xi}v_m&=& i \xi \sigma_m\bar{\lambda}, \nonumber\\
 \delta_{\xi}\lambda_\alpha&=& i\xi_\alpha \tilde{D}-ig\xi_\alpha [A,\bar{A}]
+(\sigma^{mn}\xi)_\alpha
\left( F_{mn}
+{i\over2}C_{mn}\bar{\lambda}\bar{\lambda}
\right),
\quad
 \delta_\xi\bar{\lambda}=0, \nonumber\\
 \delta_\xi \tilde{D}&=& -\xi \sigma^m D_m\bar{\lambda}
+\sqrt{2}g[\xi\psi,\bar{A}],
\nonumber\\
\delta_\xi A&=& \sqrt{2}\xi\psi,
\quad 
\delta_\xi \psi= \sqrt{2}\xi F, 
\quad
\delta_\xi F=0, 
\nonumber\\
\delta_\xi \bar{A}&=&0,
\nonumber\\
\delta_\xi\bar{\psi}&=& \sqrt{2}i \bar{\sigma}^m\xi D_m\bar{A},
\nonumber\\
\delta_\xi \bar{F}&=& i\sqrt{2}\xi\sigma^m D_m\bar{\psi}
-2gi [\bar{A},\xi\lambda]
+C^{mn}D_m\left\{ \bar{A},\xi\sigma_n\bar{\lambda}\right\}.
\end{eqnarray}
Note that the transformation of $\Phi$  is undeformed.
The deformed transformation $\delta_\eta$, which relate the gauge field 
$v_m$ to chiral fermion $\psi$, 
can be calculated in a similar way. 
But as in the analysis of $U(1)$ case, it is necessary to calculate up to
the order $O(C^2)$.
The result is
\begin{eqnarray}
 \delta_\eta v_m&=& -i \eta \sigma_m\bar{\psi}
-{\sqrt{2}\over2}C^{\alpha\beta}\eta_\alpha\left\{ \bar{A}, 
(\sigma_m\bar{\lambda})_{\beta}\right\},
\nonumber\\
 \delta_\eta \lambda^\alpha&=& \sqrt{2}\eta^\alpha \bar{F}
\nonumber\\
&&
-{\sqrt{2}\over2}C^{\alpha\beta}\eta_{\beta}\left\{ \tilde{D},\bar{A}\right\}
-{\sqrt{2}i\over2}C^{\alpha\beta}(\sigma^{mn}\eta)_\beta
\left\{ F_{mn}, \bar{A}\right\}
-{\sqrt{2}g\over2} C^{\alpha\beta}\eta_{\beta}
\left\{ \bar{A}, [\bar{A},A]\right\}
\nonumber\\
&&+{\sqrt{2}\over4}
\det C
\left(\{\bar{\lambda}\bar{\lambda},\bar{A}\}
+2\bar{\lambda}_{\dot{\alpha}}\bar{A}\bar{\lambda}^{\dot{\alpha}}\right)
\eta^{\alpha},
\nonumber\\
\delta_\eta\bar{\lambda}&=&\sqrt{2}i\bar{\sigma}^m \eta D_m\bar{A},\nonumber\\
\delta_\eta \tilde{D}&=&
-\eta \sigma^m D_m \bar{\psi}-\sqrt{2}g[\eta\lambda, \bar{A}]
-{\sqrt{2}\over2} i C^{\alpha\beta}\eta_\beta
D_m\left\{ \bar{A}, (\sigma^m\bar{\lambda})_\alpha\right\}
-ig C^{\alpha\beta}\eta_\beta
\left\{\bar{A}, [\bar{A}, \psi_\alpha]\right\} ,
\nonumber\\
\delta_\eta A&=& \sqrt{2}\eta\lambda
+ i  C^{\alpha\beta}\eta_\beta
\left\{\psi_\alpha, \bar{A}\right\} ,
\nonumber\\
\delta_\eta \psi^\alpha&=& i\eta^\alpha
\tilde{D}+ig\eta^\alpha[A,\bar{A}]
-\varepsilon^{\alpha\beta}(\sigma^{mn}\eta)_\beta F_{mn}
-i C^{\alpha\beta}\eta_{\beta}
\left\{(\bar{\lambda}\bar{\lambda})-\left\{\bar{A}, F\right\} \right\} ,
\nonumber\\
\delta_\eta F&=& i\sqrt{2}\eta\sigma^m D_m\bar{\lambda}
+2gi [\bar{A},\eta\psi] ,
\nonumber\\
\delta_\eta \bar{A}&=&0 ,
\nonumber\\
\delta_\eta\bar{\psi}_{\dot{\alpha}}&=& 
C^{\alpha\beta}\eta_\beta \sigma^m_{\alpha\dot{\alpha}}
\left\{\bar{A}, D_m\bar{A}\right\} ,
\nonumber\\
\delta_\eta \bar{F}&=& 
\sqrt{2}g
C^{\alpha\beta}\eta_\beta \left\{\bar{A}, 
[\bar{A},\lambda_\alpha]\right\} 
+
{\sqrt{2}i\over 4}\det C
\biggl[3\left\{\bar{A}, 
\left\{ \eta\sigma^{m}\bar{\lambda}, D_m\bar{A}\right\}\right\}
\nonumber\\
&&+
2 D_m\bar{A}\bar{A}\eta\sigma^m\bar{\lambda}
+
2 \eta\sigma^m\bar{\lambda}\bar{A}D_m\bar{A}
+
2\left\{\bar{A}, 
\left\{ \eta\sigma^{m}D_m\bar{\lambda}, \bar{A}\right\}\right\}
\biggr]
\label{eq:dsusyeta1}
\end{eqnarray}
Here we have used the formula $\det C=|C|^2/4$.
Note that there is an ambiguity to determine the $\delta_\eta$
transformation as noticed in the $U(1)$ case \cite{ArItOh5}.
In fact, for arbitrary functions $f_1(\bar{A})$ and $f_{2}(\bar{A})$ 
of $\bar{A}$, 
the transformation
\begin{eqnarray}
 \tilde{\delta}_\eta\lambda^{\alpha}
&=&\eta^\alpha f_1 F+F f_2 \eta^\alpha,
\nonumber\\
\tilde{\delta}_\eta\bar{F}
&=&
i(\eta\sigma^n D_n \bar{\lambda}) f_1
+i f_2 (\eta\sigma^n D_n \bar{\lambda})
+i\sqrt{2} g [\bar{A},\eta\psi]f_1+\sqrt{2}i f_2 [\bar{A},\eta\psi]
\end{eqnarray}
leaves the action invariant.
In formulas (\ref{eq:dsusyeta1}), we have chosen $f_1$ and $f_2$ such
that we recover the $U(1)$ result.
This ambiguity would be fixed if we use non(anti)commutative ${\cal
N}=2$ harmonic superspace, which will not be discussed in this  paper.

The deformed transformation $\delta_{\bar{\eta}}$ is found to be
\begin{eqnarray}
 \delta_{\bar{\eta}} v_m&=& -i \bar{\eta} \bar{\sigma}_m\psi\nonumber\\
 \delta_{\bar{\eta}} \lambda^\alpha&=& \sqrt{2}i \varepsilon^{\alpha\beta}
(\sigma^m\bar{\eta})_\beta D_m A
+i C^{\alpha\beta}\left\{\bar{\eta}\bar{\lambda},\psi_\beta\right\},
\quad
 \delta_{\bar{\eta}}\bar{\lambda}=
\sqrt{2}\bar{\eta}F,
\nonumber\\
 \delta_{\bar{\eta}} \tilde{D}&=& 
\bar{\eta} \bar{\sigma}^m D_m \psi-\sqrt{2}g[A,\bar{\eta}\bar{\lambda}],
\nonumber\\
\delta_{\bar{\eta}} A&=& 0,
\quad
\delta_{\bar{\eta}} \psi= 0,
\quad
\delta_{\bar{\eta}} F= 0,
\nonumber\\
\delta_{\bar{\eta}} \bar{A}&=&\sqrt{2}\bar{\eta}\bar{\lambda} ,
\nonumber\\
\delta_{\bar{\eta}} \bar{\psi}&=& -i\bar{\eta}\tilde{D}-ig\bar{\eta}[A,\bar{A}]
-\bar{\sigma}^{mn}\bar{\eta}F_{mn} ,
\nonumber\\
\delta_{\bar{\eta}} \bar{F}&=& 
\sqrt{2}i\bar{\eta}\bar{\sigma}^m D_m \lambda
-2gi [\bar{\eta}\bar{\psi},A]
+C^{\alpha\beta}(\sigma^m\bar{\eta})_\alpha 
D_m\left\{\psi_\beta,\bar{A}\right\}\nonumber\\
&&
-{\sqrt{2}\over4}\det C
\left\{
3\{\bar{\eta}\bar{\lambda}, \bar{\lambda}\bar{\lambda}\}
+\bar{\eta}\bar{\lambda}\bar{A}F+\bar{A}\bar{\eta}\bar{\lambda}F
-2\bar{\eta}\bar{\lambda}F\bar{A}
+2\bar{\lambda}_{\dot{\alpha}}
(\bar{\eta}\bar{\lambda})\bar{\lambda}^{\dot{\alpha}}\right\}.
\end{eqnarray}
Note that if we set $N=1$,
the cubic terms in $\bar{\lambda}$ and the commutators vanish. 
We then recover the $U(1)$ results obtained in
\cite{ArItOh5}.

We discuss some general properties obtained from the deformed 
supersymmetry.
Firstly we examine the $C$-dependence of the correlation functions.
Let us deform $C_{mn}\rightarrow C_{mn}+\delta C_{mn}$
with keeping the self-dual condition.
The variation of the Lagrangian is
\begin{eqnarray}
 \delta{\cal L}
&=&
{1\over k}
{\rm tr}
\delta C^{mn}
\Bigl\{
-{i\over2}F_{mn}\bar{\lambda}\bar{\lambda}
+{1\over4} C_{mn}(\bar{\lambda}\bar{\lambda})^2
+{i\over2} F_{mn}\left\{\bar{A}, F\right\}\nonumber\\
&&
-{\sqrt{2}\over4}\varepsilon^{\alpha\gamma}(\sigma_{mn})_{\gamma} {}^{\beta}
\left\{ D_p\bar{A}, (\sigma^p\bar{\lambda})_{\alpha}\right\}\psi_\beta
\nonumber\\
&&
-{1\over 8}C_{mn}(\bar{\lambda}\bar{\lambda})\{F,\bar{A}\}
-{1\over 4}C_{mn}\bar{A}\bar{\lambda}_{\dot{\alpha}}
F\bar{\lambda}^{\dot{\alpha}}\Bigr\} .
\end{eqnarray}
Writing $\delta_\xi=\xi^\alpha Q_\alpha$, the $Q^\alpha$ action to the
component fields are
\begin{eqnarray}
\left\{Q^\beta,\lambda_\alpha\right\}&=&
-i \delta^{\beta}_{\alpha}\tilde{D}+ig\delta^{\beta}_{\alpha}[A,\bar{A}]
+(\sigma^{mn})_{\alpha} {}^\beta
\left( F_{mn}+{i\over2} C_{mn} (\bar{\lambda}\bar{\lambda})\right),
\nonumber\\
\left\{Q^\beta,\psi_\alpha\right\}&=& -\sqrt{2}\delta^{\beta}_{\alpha} F ,
\nonumber\\
\left\{Q^\beta,\bar{\psi}_{\dot{\alpha}}\right\}&=&-i\sqrt{2}D_m\bar{A}
\varepsilon^{\beta\gamma}\sigma^m_{\gamma\dot{\alpha}} .
\end{eqnarray}
Then we find that $\delta{\cal L}$ is $Q$-exact, 
$ \delta{\cal L}=\left\{Q^{\alpha}, \Lambda_\alpha\right\}$, where 
\begin{eqnarray}
\Lambda_\alpha&=&
{\rm tr}\delta C^{mn}\Bigl\{
{i\over4}
(\sigma_{mn})_{\alpha}{}^{\beta}\lambda_{\beta}
\left(\bar{\lambda}\bar{\lambda}-\left\{\bar{A},F\right\}\right)
-{i\over4}
(\sigma_{mn})_{\alpha} {}^{\beta}
(\bar{\psi}\bar{\lambda}+\bar{\lambda}\bar{\psi})
\psi_\beta
\nonumber\\
&&
-{1\over \sqrt{2}} 
\Bigl\{
{1\over8} C_{mn}\{\psi_\alpha,\bar{A}\}(\bar{\lambda}\bar{\lambda})
+{1\over4} C_{mn}\bar{\lambda}_{\dot{\alpha}}
\bar{A}\bar{\lambda}^{\dot{\alpha}}\psi_{\alpha}
\Bigr\}\Bigr\}.
\end{eqnarray}
In the case of non(anti)commutative ${\cal N}=1$ super
Yang-Mills theory\cite{Im}, $\delta {\cal L}$ is shown to be
$Q$-exact, which means that the correlator $\langle {\cal O}_1\cdots 
{\cal O}_n\rangle$ for $Q$-invariant operators ${\cal O}_i$ is $C$-independent.
The antichiral gluino condensates $\langle
{\rm tr}\bar{\lambda}\bar{\lambda}\rangle$, in particular,
 does not  have $C$-correction.
In the ${\cal N}=2$ case, 
the correlator $\langle {\cal O}_1\cdots {\cal O}_n\rangle$ is also 
$C$-independent if the vacuum is $Q$-invariant:
$Q|0\rangle=\langle 0|Q=0$ \cite{Im}.
For example, the antichiral scalar field 
$\bar{A}$ is a $Q$-invariant operator.  Therefore,
the vacuum expectation value 
$\langle {\rm tr}\bar{A}^2\rangle$ is $C$-independent.
In the low-energy effective theory, this can be expressed in terms of 
anti-holomorphic prepotential $\bar{{\cal F}}(\bar{a})$\cite{EgYa} such as 
\begin{equation}
\langle {\rm tr}\bar{A}^2\rangle \sim 
\sum_{i}
 \bar{a}_{i} \frac{\partial \bar{{\cal F}}(\bar{a})}{\partial \bar{a}_{i}}
-2\bar{{\cal F}}(\bar{a}), \label{prep}
\end{equation}
where $\bar{a}_{i}$ is the vacuum expectation value 
of the low-energy $U(1)^N$ Higgs fields.
Since $\bar{{\cal F}}(\bar{a})$ is expanded in the QCD scale parameter,
(\ref{prep}) implies that the instanton corrections to the 
anti-holomorphic prepotential are not $C$-deformed.
On the other hand, the holomorphic prepotential would have
$C$-corrections.

We next study the (constrained) instanton equations of the 
deformed ${\cal N}=2$ supersymmetric $U(N)$ gauge theory. 
Since the deformed part of the gauge field in the action is the same as
that of the ${\cal N}=1$ theory\cite{Im}, 
the deformed instanton equation for the gauge field is $F^{-}_{mn}=0$ 
in the self-dual case
and $F^{+}_{mn}+\frac{i}{2}C_{mn}\bar{\lambda}\bar{\lambda}=0$ 
in the anti-self-dual case\cite{Im,inst}. 
Here $F^{\pm}_{mn}=\frac{1}{2}(F_{mn}\pm\tilde{F}_{mn})$. 
The other fields must satisfy the equation of motion in the instanton 
background.

We begin with considering the case
where the vacuum expectation values of 
$A$ and $\bar{A}$ are zero.
In this case we expect that the solutions are exact. 
{}From the Lagrangian (\ref{eq:lag1}), we obtain the equations of motion 
\begin{eqnarray}
&&D_{m}F^{mn}-g\{\bar{\lambda},\bar{\sigma}^{n}\lambda\}
-g\{\bar{\psi},\bar{\sigma}^{n}\psi\}
-ig[\bar{A},D^{n}A]+ig[D^{n}\bar{A},A]
\nonumber\\
&& +iD_{m}(C^{mn}\bar{\lambda}\bar{\lambda})
+\frac{i}{\sqrt{2}}g\Bigl[\bar{A},
C^{\alpha\beta}[(\sigma^{n}\bar{\lambda})_{\alpha},\psi_{\beta}]
\Bigr]=0,\nonumber\\
&&
\bar{\sigma}^{m}D_{m}\lambda+\sqrt{2}g[A,\bar{\psi}]+C^{mn}(F_{mn}
+\frac{i}{2}C_{mn}\bar{\lambda}\bar{\lambda})\bar{\lambda}=0,\nonumber\\
&&\sigma^{m}D_{m}\bar{\lambda}+\sqrt{2}g[\bar{A},\psi]=0,\quad
\bar{\sigma}^{m}D_{m}\psi-\sqrt{2}g[A,\bar{\lambda}]=0,\nonumber\\
&&(D_{m}\bar{\psi}\bar{\sigma}^{m})^{\alpha}
+\sqrt{2}g[\bar{A},\lambda^{\alpha}]
+\frac{i}{\sqrt{2}}C^{\alpha\beta}\{D_{m}
\bar{A},(\sigma^{m}\bar{\lambda})_{\beta}\}=0,\nonumber\\
&&D^{2}A-i\sqrt{2}g\{\psi,\lambda\}-g^2\Bigl[[A,\bar{A}],A\Bigr]
+\frac{1}{\sqrt{2}}C^{\alpha\beta}
D_{m}[(\sigma^{m}\bar{\lambda})_{\alpha},\psi_{\beta}]=0,\nonumber\\
&&
D^{2}\bar{A}-i\sqrt{2}g\{\bar{\psi},\bar{\lambda}\}
-g^2\Bigl[[A,\bar{A}],\bar{A}\Bigr]
=0,\label{eom}
\end{eqnarray}
where the auxiliary fields $F$, $\bar{F}$ and $\tilde{D}$ have been
eliminated.
In the anti-self-dual background, 
if we set $\psi=\lambda=0$, the equation of motion for $A$ becomes 
$D^{2}A-g^2\Bigl[[A,\bar{A}],A\Bigr]=0$, from which $A=0$ is
an exact solution as in the undeformed case. 
From $A=0$ and absence of chiral zero-modes in the anti-self-dual background, 
$\psi=\lambda=0$ is shown to be also an exact solution of the equations of 
motion for $\psi$ and $\lambda$. By a similar argument, 
$\bar{\lambda}=\bar{\psi}=0$ and $\bar{A}=0$ are an exact solution 
of the equation of motion in the self-dual background.
Then the first equation of (\ref{eom}) is satisfied. 
In this paper we discuss the case of the gauge group $U(2)$ since 
the structure of the equations of motion (\ref{eom}) becomes simple
as in \cite{Im,inst}.  
We decompose  $U(2)$ into the 
$SU(2)$ and the $U(1)$ parts: $v_{m}=v_{SU(2),m}+v_{U(1),m}$ where
\begin{eqnarray}
v_{SU(2),m}&=&v^{a}_{m}t^{a}, \quad (a=1,2,3), \quad
v_{U(1),m}=v^{4}_{m}t^{4},
\end{eqnarray}
and $t^a=\tau^a$ and $t^4=1$. $\tau^a$ are the Pauli matrices.
Other fields can be decomposed similarly.
The equations of motion of 
$\bar{\lambda}_{U(1)}$, $\psi_{U(1)}$ and $\bar{A}_{U(1)}$ 
become
\begin{equation}
\sigma^{m}\partial_{m}\bar{\lambda}_{U(1)}=
\bar{\sigma}^{m}\partial_{m}\psi_{U(1)}=
\partial^{2}\bar{A}_{U(1)}=0,\label{u11}
\end{equation}
Since eqs. (\ref{u11}) do not have any localized solution
in four-dimensional Euclidean space, 
 we have
\begin{equation}
\bar{\lambda}_{U(1)}=\psi_{U(1)}=0, \quad \bar{A}_{U(1)}=const.\label{trivial}
\end{equation}
Using (\ref{trivial}), one finds that
only the equation of motion of $\lambda_{SU(2)}$ is deformed:
\begin{equation}
\bar{\sigma}^{m}D_{m}\lambda_{SU(2)}+\sqrt{2}g[A_{SU(2)},\bar{\psi}_{SU(2)}]
+\frac{1}{2}C^{mn}[F_{SU(2),mn},\bar{\lambda}_{SU(2)}]=0,\label{su21}
\end{equation}
while the equations of motion of the other $SU(2)$ fields are not deformed.
Therefore we find that the instanton equations for the $SU(2)$ gauge field
are not deformed.
In the case where the vacuum expectation values of $A$
and $\bar{A}$ are zero, we obtain the exact self-dual solution
\begin{eqnarray}
&&
v_{SU(2),m}=g^{-1}v_{m}^{(0)},\quad
\lambda_{SU(2)}=g^{-1/2}\lambda^{(0)},
\quad
\psi_{SU(2)}=g^{-1/2}\psi^{(0)},\nonumber\\
&&
A_{SU(2)}=A^{(0)},\quad 
\bar{\lambda}_{SU(2)}=\bar{\psi}_{SU(2)}=\bar{A}_{SU(2)}=0,\label{SD}
\end{eqnarray}
which is not deformed by $C$.
Here $\lambda^{(0)}$ and $\psi^{(0)}$ are the zero-modes in the
self-dual instanton background. $A^{(0)}$ satisfies the equation
$D^{2}A^{(0)}-i\sqrt{2}\{\psi^{(0)},\lambda^{(0)}\}=0$.
Since $\bar{\lambda}_{SU(2)}=0$, the $C$-dependent term in eq. (\ref{su21}) 
vanishes and 
$\lambda_{SU(2)}$ is the same as the solution of the undeformed theory. 
The anti-self-dual solution for the $SU(2)$ part becomes
\begin{eqnarray}
&&
v_{SU(2),m}=g^{-1}v_{m}^{(0)},\quad
\bar{\lambda}_{SU(2)}=g^{-1/2}\bar{\lambda}^{(0)},
\quad
\bar{\psi}_{SU(2)}=g^{-1/2}\bar{\psi}^{(0)},\nonumber\\
&&
\bar{A}_{SU(2)}=\bar{A}^{(0)}, \quad 
\lambda_{SU(2)}=\psi_{SU(2)}=A_{SU(2)}=0,\label{ASD}
\end{eqnarray}
which are not also $C$-deformed.\footnote{
In the case of singlet deformation, it is shown in \cite{IvZu} 
that the anti-self-dual solution is not deformed by $C$.
}
Here $\bar{\lambda}^{(0)}$ and $\bar{\psi}^{(0)}$ are the zero-modes in the
anti-self-dual instanton background. $\bar{A}^{(0)}$ fulfills 
$D^{2}\bar{A}^{(0)}-i\sqrt{2}\{\bar{\psi}^{(0)},\bar{\lambda}^{(0)}\}=0$.
Since $v_{SU(2),m}=g^{-1}v_{m}^{(0)}$ is anti-self-dual, 
we get $C^{mn}F_{SU(2),mn}=0$. Therefore we find that 
(\ref{su21}) has the solution $\lambda_{SU(2)}=0$.
From eq. (\ref{trivial}), 
the equations of motion of other $U(1)$ fields become 
\begin{eqnarray}
&&\partial^{m}\left(F^{4}_{mn}
+iC_{mn}\bar{\lambda}^{a}\bar{\lambda}^{a}\right)=0,\label{u12}\\
&&\bar{\sigma}^{m}\partial_{m}\lambda^{4}
+C^{mn}F^{a}_{mn}\bar{\lambda}^{a}=0,
\label{u13}\\
&&\left(\partial_{m}\bar{\psi}^{4}\bar{\sigma}^{m}\right)^{\alpha}
+i\sqrt{2}C^{\alpha\beta}
\partial_{m}\left(\bar{A}^{a}\sigma^{m}\bar{\lambda}^{a}\right)_{\beta}
=0,\label{u14}\\
&&\partial^{2}A^{4}+\sqrt{2}C^{\alpha\beta}
\partial_{m}\left\{(\sigma^{m}\bar{\lambda}^{a})_{\alpha}
\psi^{a}_{\beta}\right\}=0.\label{u15}
\end{eqnarray}
Eqs. (\ref{u12})--(\ref{u15}) are regarded as 
the free field equations with the source made of the $SU(2)$ fields. 
The solution of (\ref{u12}) in the Lorentz gauge is obtained in \cite{inst} as
$v^{4}_{m}=iC_{mn}\partial^{n}\Psi$, where $\Psi$ obeys 
$\partial^{2}\Psi=\bar{\lambda}^{a}\bar{\lambda}^{a}$. 
In the self-dual background (\ref{SD}), 
eqs. (\ref{u12})--(\ref{u15}) have only trivial solutions 
$F_{mn}^{4}=\lambda^{4}=\bar{\psi}^{4}=0,\ A^{4}=const.$
since $\bar{\lambda}^{a}=0$. 
As a consequence, the self-dual instanton equation $F^{4,-}_{mn}=0$ is 
satisfied. 
In the anti-self-dual background (\ref{ASD}), 
$v^{4}_{m}$ satisfies the deformed anti-self-dual instanton equation 
$F^{4,+}_{mn}+\frac{i}{2}C_{mn}\bar{\lambda}^{a}\bar{\lambda}^{a}=0$.
Eqs. (\ref{u13}) and (\ref{u15}) have trivial solutions 
$\lambda^{4}=0,\ A^{4}=const.$ due to $C^{mn}F^{a}_{mn}=\psi^{a}=0$. 
On the other hand, from eq. (\ref{u14}), 
$\bar{\psi}^{4}$ depends on $C$.

We next consider the case where the vacuum expectation values of $A$ 
and $\bar{A}$ are nonzero, which corresponds to the constrained
instanton
solutions (see \cite{instreview} for a review).
When the $SU(2)$ instantons 
are self-dual,
we expand the fields in the coupling constant $g$:
\begin{eqnarray}
v_{SU(2),m}&=&g^{-1}v_{m}^{(0)}+gv_{m}^{(1)}+\cdots,\quad
F_{SU(2),mn}=g^{-1}F_{mn}^{(0)}+gF_{mn}^{(1)}+\cdots,\nonumber\\
\lambda_{SU(2)}&=&g^{-1/2}\lambda^{(0)}+g^{3/2}\lambda^{(1)}+\cdots,\quad
\bar{\lambda}_{SU(2)}=g^{1/2}\bar{\lambda}^{(0)}+g^{5/2}\bar{\lambda}^{(1)}
+\cdots,\nonumber\\
\psi_{SU(2)}&=&g^{-1/2}\psi^{(0)}+g^{3/2}\psi^{(1)}+\cdots,\quad
\bar{\psi}_{SU(2)}=g^{1/2}\bar{\psi}^{(0)}+g^{5/2}\bar{\psi}^{(1)}+\cdots,
\nonumber\\
A_{SU(2)}&=&g^{0}A^{(0)}+g^{2}A^{(1)}+\cdots,\quad
\bar{A}_{SU(2)}=g^{0}\bar{A}^{(0)}+g^{2}\bar{A}^{(1)}+\cdots,
\end{eqnarray}
where $v_{m}^{(0)}$ is self-dual $F_{mn}^{(0)}=\tilde{F}_{mn}^{(0)}$.
We are interested in the leading terms of the $SU(2)$ fields. 
The equations of motion of 
the fields are 
\begin{eqnarray}
&&\bar{\sigma}^{m}D_{m}\lambda^{(0)}
+\frac{1}{2}C^{mn}[F^{(0)}_{mn},\bar{\lambda}^{(0)}]=0,
\label{eom2a}\\
&&
\sigma^{m}D_{m}\bar{\lambda}^{(0)}+\sqrt{2}[\bar{A}^{(0)},\psi^{(0)}]=0,\quad 
\bar{\sigma}^{m}D_{m}\psi^{(0)}=0,\quad D^{2}\bar{A}^{(0)}=0,
\label{eom4a}\\
&&
\sigma^{m}D_{m}\bar{\psi}^{(0)}-\sqrt{2}[\bar{A}^{(0)},\lambda^{(0)}]=0,\quad 
D^{2}A^{(0)}-i\sqrt{2}\{\psi^{(0)},\lambda^{(0)}\}=0.
\label{eom6a}
\end{eqnarray}
Only the eq. (\ref{eom2a}) is deformed by $C$. 
For example, the $C$-deformed supersymmetric zero-mode 
can be 
obtained from the supersymmetry transformation (\ref{eq:dsusyeta1}) as 
\begin{equation}
\lambda^{(0),\alpha}=-\sqrt{2}i(\bar{A}_{U(1)}
C\eta)^{\beta}(\sigma^{mn})_{\beta}^{\ \alpha}
F_{mn}^{(0)}.\label{sol1}
\end{equation}
We note that $\bar{\psi}^{(0)}$ and $A^{(0)}$ are also deformed 
since eqs. (\ref{eom6a}) contain 
$\lambda^{(0)}$ as a source.

In the case of anti-self-dual background $F_{mn}^{(0)}=-\tilde{F}_{mn}^{(0)}$, 
the $g$-expansion of the $SU(2)$ fields becomes 
\begin{eqnarray}
v_{SU(2),m}&=&g^{-1}v_{m}^{(0)}+gv_{m}^{(1)}+\cdots,\quad
F_{SU(2),mn}=g^{-1}F_{mn}^{(0)}+gF_{mn}^{(1)}+\cdots,\nonumber\\
\lambda_{SU(2)}&=&g^{1/2}\lambda^{(0)}+g^{5/2}\lambda^{(1)}+\cdots,\quad
\bar{\lambda}_{SU(2)}=g^{-1/2}\bar{\lambda}^{(0)}+g^{3/2}\bar{\lambda}^{(1)}+
\cdots,\nonumber\\
\psi_{SU(2)}&=&g^{1/2}\psi^{(0)}+g^{5/2}\psi^{(1)}+\cdots,\quad
\bar{\psi}_{SU(2)}=g^{-1/2}\bar{\psi}^{(0)}+g^{3/2}\bar{\psi}^{(1)}+\cdots,
\nonumber\\
A_{SU(2)}&=&g^{0}A^{(0)}+g^{2}A^{(1)}+\cdots,\quad
\bar{A}_{SU(2)}=g^{0}\bar{A}^{(0)}+g^{2}\bar{A}^{(1)}+\cdots.
\end{eqnarray}
The leading terms of the equations of motion are 
\begin{eqnarray}
&&
\bar{\sigma}^{m}D_{m}\lambda^{(0)}+\sqrt{2}[A^{(0)},\bar{\psi}^{(0)}]
+\frac{1}{2}C^{mn}[F^{(1)}_{mn},\bar{\lambda}^{(0)}]=0,\label{eom2b}\\
&&
\sigma^{m}D_{m}\bar{\lambda}^{(0)}=0,\quad 
\bar{\sigma}^{m}D_{m}\psi^{(0)}-\sqrt{2}[A^{(0)},\bar{\lambda}^{(0)}]=0,
\label{eom4b}\\
&&
\sigma^{m}D_{m}\bar{\psi}^{(0)}=0,\quad 
D^{2}A^{(0)}=0,\quad 
D^{2}\bar{A}^{(0)}-i\sqrt{2}\{\bar{\psi}^{(0)},\bar{\lambda}^{(0)}\}=0.
\label{eom7b}
\end{eqnarray}
Eqs. (\ref{eom4b}) and (\ref{eom7b}) can be solved as in the 
undeformed theory. 
But the $C$-deformed source
in $\lambda^{(0)}$ comes  from the 
order $g$ contributions of the gauge fields.
At the leading order in $g$, the $SU(2)$ anti-self-dual solutions are not
modified by the deformation parameter $C$.
This is consistent with the result obtained from the $Q$-exactness of
the $C$-variation of the action.
In the $U(1)$ part, the equations of motions 
(\ref{u12})--(\ref{u15}) would have $C$-dependent solutions. 
A detailed analysis of a instanton calculus will be discussed 
in a separate paper.

\subsection*{Acknowledgements}
H. N. is supported by Grant-in-Aid for Scientific Research 
from the Ministry of Education, Culture, Sports, Science and 
Technology, Japan No. 16028203 for the priority area 
``origin of mass".


\begin{thebibliography}{99}

\bibitem{ScNi}
J.~H.~Schwarz and P.~Van Nieuwenhuizen,
Lett.\ Nuovo Cim.\  {\bf 34}, 21 (1982). 


 \bibitem{Se} N.~Seiberg,
JHEP {\bf 0306}, 010 (2003),
hep-th/0305248. 


 \bibitem{OoVa} H.~Ooguri and C.~Vafa,
Adv. Theor. Math. Phys. {\bf 7} (2003) 53,
hep-th/0302109; 
Adv. Theor. Math. Phys. {\bf 7} (2004) 405, hep-th/0303063. 

\bibitem{BeSe}
N.~Berkovits and N.~Seiberg,
JHEP {\bf 0307} (2003) 010,
hep-th/0306226. 


 \bibitem{DeGrNi} J.~de Boer, P.~A.~Grassi and P.~van Nieuwenhuizen,
Phys. Lett. {\bf B574} (2003) 98,
hep-th/0302078. 


\bibitem{Ne}
N.A.~Nekrasov, Adv. Theor. Math. Phys. {\bf 7} (2004) 831, hep-th/0206161.

\bibitem{DiVa}
R.~Dijkgraaf and C.~Vafa, Nucl. Phys. {\bf B644} (2002) 3, hep-th/0206255.

\bibitem{KlPeTa}
D.~Klemm, S.~Penati and L.~Tamassia,
Class.\ Quant.\ Grav.\  {\bf 20} (2003) 2905,
hep-th/0104190; \\ 
S.~Ferrara and M.~A.~Lledo,
JHEP {\bf 0005} (2000) 008,
hep-th/0002084; \\
S.~Ferrara, M.~A.~Lledo and O.~Macia,
JHEP {\bf 0309} (2003) 068,
hep-th/0307039.

\bibitem{FeSo}
S.~Ferrara and E.~Sokatchev, 
Phys. Lett. {\bf B579} (2004) 226,
hep-th/0308021.

\bibitem{IvLeZu}
E. Ivanov, O. Lechtenfeld and B. Zupnik,
JHEP {\bf 0402} (2004) 012,
hep-th/0308012;
Nucl.\ Phys.\ B {\bf 707} (2005) 69, 
hep-th/0408146. 

\bibitem{ArItOh2}
T.~Araki, K.~Ito and A.~Ohtsuka, 
JHEP {\bf 0401} (2004) 046, 
hep-th/0401012.


\bibitem{KeSa}
S.V.~Ketov and S.~Sasaki,
Phys.\ Lett.\ B {\bf 595} (2004) 530,
hep-th/0404119;
Phys.\ Lett.\ B {\bf 597} (2004) 105,
hep-th/0405278;
Int.\ J.\ Mod.\ Phys.\ A {\bf 20} (2005) 4021,
hep-th/0407211.

\bibitem{IvZu}
E.~A.~Ivanov and B.~M.~Zupnik,
Theor.\ Math.\ Phys.\  {\bf 142} (2005) 197,
hep-th/0405185.


\bibitem{FeIvLeSoZu}
S.~Ferrara, E.~Ivanov, O.~Lechtenfeld, E.~Sokatchev and B.~Zupnik,
Nucl.\ Phys.\ B {\bf 704} (2005) 154, 
hep-th/0405049.


\bibitem{ArItOh4}
T.~Araki, K.~Ito and A.~Ohtsuka, Phys. Lett. {\bf B606} (2005) 202,
hep-th/0410203. 

\bibitem{ArItOh5}
T.~Araki, K.~Ito and A.~Ohtsuka, 
JHEP {\bf 0505} (2005) 074,
hep-th/0503224.



\bibitem{ArIt3}
T.~Araki and K.~Ito, 
Phys. Lett. {\bf B595} (2004) 513, 
hep-th/0404250.

\bibitem{ArItOh1}
T.~Araki, K.~Ito and A.~Ohtsuka, 
Phys. Lett. {\bf B573} (2003) 209, 
hep-th/0307076. 




\bibitem{Im}
A.~Imaanpur, 
JHEP {\bf 0309} (2003) 077, hep-th/0308171;
JHEP {\bf 0312} (2003) 009, 
hep-th/0311137.


\bibitem{inst}
P.~A.~Grassi, R.~Ricci and D.~Robles-Llana, 
JHEP {\bf 0407} (2004) 065,
hep-th/0311155;\\
R.~Britto, B.~Feng, O.~Lunin and S.~J.~Rey, 
Phys.\ Rev.\ D {\bf 69} (2004) 126004,
hep-th/0311275;\\
M.~Bill\'o, M.~Frau, I.~Pesando and A.~Lerda,
JHEP {\bf 0405} (2004) 023,
hep-th/0402160;\\
M.~Bill\'o, M.~Frau, F.~Lonegro and A.~Lerda,
JHEP {\bf 0505} (2005) 047,
hep-th/0502084;\\
S.~Giombi, R.~Ricci, D.~Robles-Llana and D.~Trancanelli,
hep-th/0505077.


\bibitem{WeBa}
J.~Wess and J.~Bagger, ``Supersymmetry and
Supergravity,''
Princeton University Press, 1992.


\bibitem{GaIvOgSo}
A.~Galperin, E.~Ivanov, V.~Ogievetsky and E.~Sokatchev,
``Harmonic Superspace'', Cambridge University Press , 2001.


\bibitem{EgYa}
J.~Sonnenschein, S.Theisen and S.~Yankielowicz, Phys. Lett. {\bf B367} 
(1996) 145, hep-th/9510129;\\
T.~Eguchi and S.-K.~Yang, Mod. Phys. Lett. {\bf A11} (1996) 131, 
hep-th/9510183.


\bibitem{instreview}
N.~Dorey, T.J.~Hollowood, V.V.~Khoze and M.P.~Mattis, 
Phys. Rept. {\bf 371} (2002) 231, hep-th/0206063





\end{thebibliography}
\end{document}